\documentclass[
    ,final            
  ]
  {aipproc}

\layoutstyle{6x9}

\begin{document}

\title{Hadronic spectra in AdS / QCD correspondence}


\author{Alfredo Vega and Ivan Schmidt}{
  address={Departamento de F\'\i sica y Centro de Estudios
Subat\'omicos,\\  Universidad T\'ecnica Federico Santa Mar\'\i a,
\\ Casilla 110-V, Valpara\'\i so, Chile}
}



\begin{abstract}
We present an holographical soft wall model which is able to
reproduce Regge spectra for hadrons with an arbitrary number
of constituents. The model includes the anomalous dimension of
operators that create hadrons, together with a dilaton, whose form
is suggested by Einstein's equations and the AdS metric.
\end{abstract}

\maketitle


\section{Introduction}

From its beginnings, progress in QCD at low energies has been
impeded because there are no good analytical tools available in order to work
with strongly coupled Yang Mills theories. Nevertheless, in the last
few years the AdS / CFT ideas has provided a new approach that could
improve this situation.

At present a dual to QCD is unknown, but a simple approach known as
Bottom - Up has been quite successful in several concrete QCD applications, such as
in hadronic scattering processes \cite{Scattering}, hadronic spectra \cite{Spectra, Forkel, VegaSchmidt1, VegaSchmidt2}, hadronic couplings and chiral symmetry breaking \cite{Chiral}, mesonic wave function \cite{WaveFunction}, among other applications.

Here we summarize the main ideas developed in \cite{VegaSchmidt1, VegaSchmidt2}, where a soft wall holographical model that describes hadronic spectra  with an arbitrary number of constituents was proposed.

The present work has been structured as follow. Section II is a summary of the model considered. In section III
we give some examples, and finally in IV we present some conclusions.

\section{Spectra of hadrons in AdS / QCD.}

We begin by considering an asymptotically AdS space defined by
\begin{equation}
 \label{metrica}
 ds^2 = e^{2 A(z)} ( \eta_{\mu\nu} dx^{\mu} dx^{\nu} ) ,
\end{equation}
and an action for arbitrary spin modes (which depends on the spin of the hadron described).

\begin{figure}[h]
  \begin{tabular}{cc}
    \includegraphics[width=2.0in]{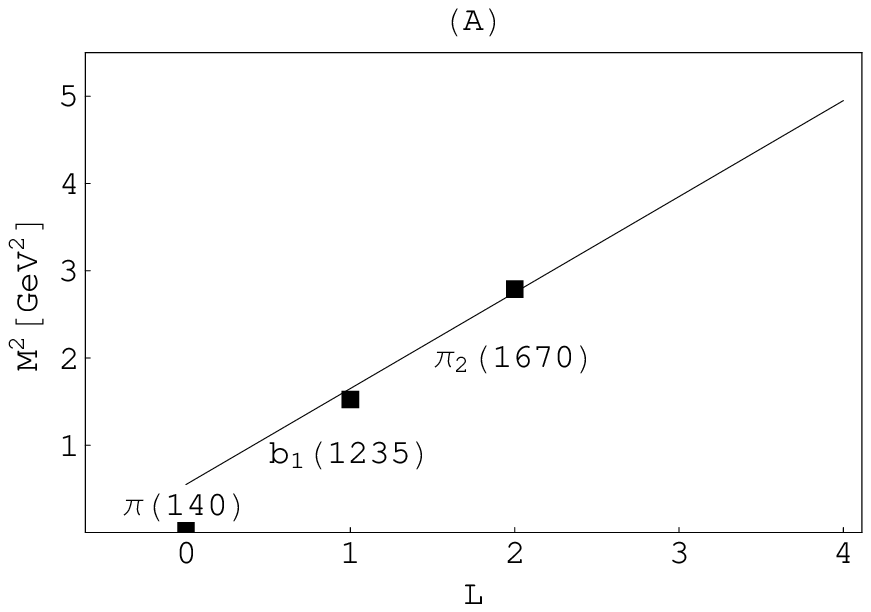} &
    \includegraphics[width=2.0in]{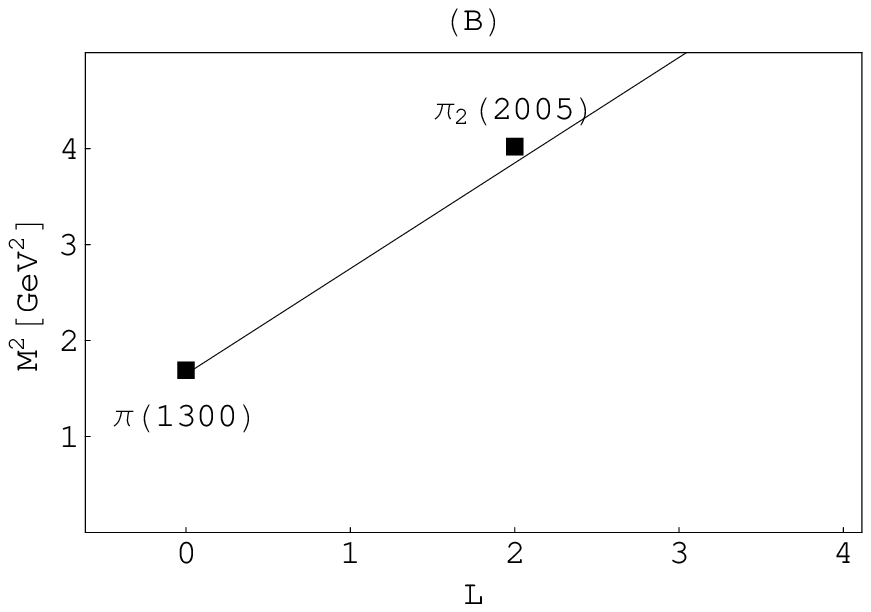}
  \end{tabular}
\caption{Spectra of scalar mesons, calculated within the Soft Wall
model. The figures correspond to different radial excitations. (A)
$n=0$ (B) $n=1$ . } 
\end{figure}
For modes with integer arbitrary spin the corresponding equation of motion is \cite{VegaSchmidt2}
\begin{equation}
 \label{EcuacionBosonGeneral}
 \partial_{z}^{2} \varphi - [\partial_{z} (\Phi (z) - \beta A (z))] \partial_{z} \varphi + [ M^{2} - m_{5}^{2} e^{2 A(z)} ] \varphi = 0 ,
\end{equation}
where $\beta = k (2 S - 1)$, k is a constant and S corresponds to the spin of the mode
considered and $\Phi (z)$ is a dilaton field.  We fix $\beta$ for each spin mode using experimental data.

Consider an AdS space, i.e. $e^{2 A(z)} = \frac{R^{2}}{z^{2}}$, with a quadratic dilaton ($\Phi (z) = \kappa^{2} z^{2}$, and with $\kappa = \textbf{constant}$).  This allows us to obtain spectra with Regge behavior.

Nevertheless, according to \cite{Kirsch}, in the second order Dirac equation the dilaton in AdS can be factorized, and the equation looks like

\begin{equation}
 \label{EcuacionDirac}
 \partial_{z}^{2} f_{\pm} - \frac{4}{z} \partial_{z} f_{\pm} + \biggl[ M^{2} + \frac{6}{z^{2}} - \frac{m_{5}^{2} R^{2}}{z^{2}} + \frac{\gamma m_{5} R}{z^{2}} \biggr] f_{\pm} = 0 ,
\end{equation}
where $\gamma = \pm 1$, depending on whether we are considering the
left or right part.

When $m_{5} R$ is constant, the spectra obtained doesn't have Regge behavior.

Since in this case the dilaton field cannot improve the situation with respect to hard wall models, it is necessary to try other possibilities, for example a trivial dilaton but with a family of metrics \cite{Forkel, DePaula}, or consider an AdS metric with different dilaton and include anomalous dimension for the operators that create the hadrons, as considered in \cite{VegaSchmidt2}.

Einstein's equations determine the dilaton directly from the metric ($\Phi' = \sqrt{3 A'^{2} - 3 A''}$)\cite{DePaula}. So, our model is defined by $A(z) = \rho \ln ( R / z)$, then $\Phi(z) = \lambda ln(z)$, where $\lambda$ depends on $\rho$, although in order to get equations with exact solutions, we will use $\lambda = 2$.

The AdS / CFT dictionary tell us that the twist dimension of operators on the CFT side and the conformal dimension of the AdS modes must be equal, and this establish possible values for $m_{5}^{2} R^{2}$, that correspond to \cite{VegaSchmidt2}
\[ m_{5}^{2} R^{2} = \left\lbrace
  \begin{array}{c l}
    (\Delta_{0} + L - S + \omega z^{2})(\Delta_{0} + L - S - 3 + \beta + \omega z^{2}) & \texttt{;Integer spin}\\
    \Delta_{0} + L - S - 3 + \omega z^{2} & \texttt{;Spin 1/2}.
  \end{array}
\right. \]
where in each case an anomalous dimension was considered for the operators.
\begin{figure}[h]
  \begin{tabular}{cc}
    \includegraphics[width=2.0in]{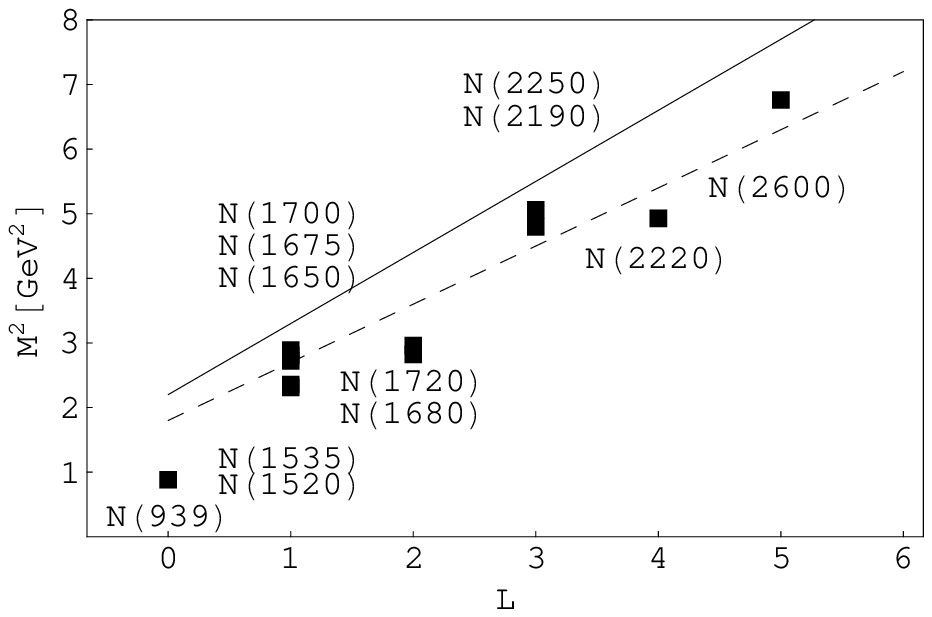} &
    \includegraphics[width=2.0in]{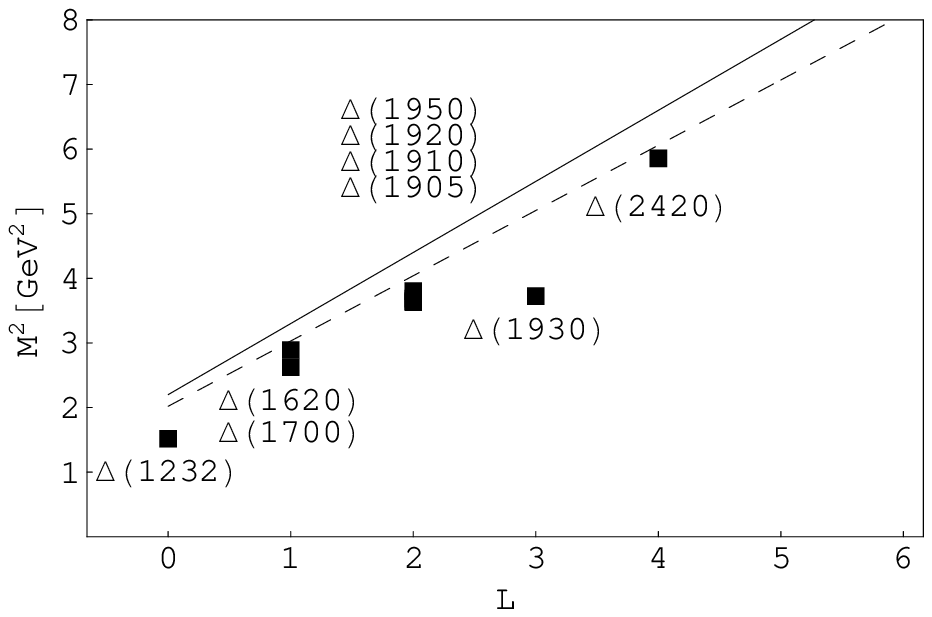}
  \end{tabular}
\caption{Nucleons and $\Delta$ resonances spectra. The continuous
line is the model prediction using an  universal value of $A = 1.1
GeV^{2}$, while the dashed line was obtained using Regge slopes
adjusted to each case, with values $A = 0.9 GeV^{2}$ for nucleons
and $A = 1.01 GeV^{2}$ for $\Delta$ resonances \cite{Forkel}.}
\end{figure}

\section{Some hadronic spectrum.}

The spectrum for all cases considered, is \cite{VegaSchmidt2}
\begin{equation}
 \label{Masas}
 M^{2} = A [n + L + v],
\end{equation}
where $A = 4 \omega$ is the Regge slope, and $v$ is given by
\[ v = \left\lbrace
  \begin{array}{c l}
    \Delta_{0} + \frac{\beta}{2} - 1 - S & ; \texttt{If S is integer}\\
    \Delta_{0} - \frac{5}{2} & ; \texttt{spin 1/2}.
  \end{array}
\right. \]

Notice that the model gives us the Regge slope in terms of $\omega$, which constitutes a phenomenological input in our model. Equation (\ref{Masas}) can be applied to different kinds of hadrons with an arbitrary number of constituents, considering different values for $\Delta_{0}$ and the right value for $\beta$ depending on the spin, and in general we take $A \sim 1.1 GeV^{2}$, which can be considered approximately universal for all trajectories \cite{Iachello}.

The spectrum for some scalar mesons is shown in Fig 1, while some
examples about model predictions for scalar exotic hadrons appear in
Table 1.

In the scalar case a universal Regge slope, with value 1.1$Gev^{2}$ and $\beta = -3$, was used.  $\Delta_0$ was calculated considering that each quarks and / or antiquarks contribute with 3/2 to $\Delta_{0}$, and that gluons contribute
with 2 to $\Delta_{0}$.
\begin{table}[h]
\caption{Scalar exotic hadron masses, with $n = L = 0$. We consider
hadrons with n quarks (and / or antiquarks) and m gluons.}
\begin{tabular}{ c c c | c c c | c c c }
  \hline
  & $\Delta_0$ & & & (nQ)(mG) & & & M [GeV] & \\
  \hline
  & 4 & & & (2G) & & & 1.28 & \\
  & 5 & & & (2Q)(1G) & & & 1.66 & \\
  & 6 & & & (4Q) & & & 1.96 & \\
  \hline
\end{tabular}
\end{table}

For other integer spin hadrons it is necessary to do a similar analysis, but take s different value for $\beta$. For example, in the vector case $\beta = -1$ was used in \cite{VegaSchmidt2}.

For the spin 1/2 case, as one can see from Fig 2, using an universal Regge slope gives
results somewhat higher than the experimental data, but using a value of
0.9 $[GeV^{2}]$ \cite{Forkel}, adjusted to baryonic data, the
results are better. Both values are used in Table 2, where model predictions for some spin 1/2 exotic hadrons are shown.
\begin{table}[h]
\caption{Spin 1/2 exotic hadron masses with $n = L = 0$. We consider
hadrons with n quarks (and / or antiquarks) and m gluons. Column
$M_{U}$ was calculated using $A = 1.1 GeV^{2}$, the universal Regge
slope used in this work, while M contains the results obtained using
$A = 0.9 GeV^{2}$, a value fixed from nucleon data \cite{Forkel}.}
\begin{tabular}{ c c c | c c c | c c c | c c c }
  \hline
  & $\Delta_0$ & & & (nQ)(mG) & & & $M_{U}$ [GeV] & & & M [GeV] & \\
  \hline
  & 13/2 & & & (1Q)(3G) & & & 2.10 & & & 2.01 & \\
  & 15/2 & & & (5Q) & & & 2.35 & & & 2.25 & \\
  & 17/2 & & & (3Q)(2G) & & & 2.57 & & & 2.46 & \\
  \hline
\end{tabular}
\end{table}
On the other side, solutions to Rarita - Schwinger equation in AdS space are more
difficult to get, but its spectrum is similar to the Dirac case, as you can see for example in
Ref. \cite{Forkel}. As is possible to see in Fig 2, again the
results are somewhat high, but using $A = 1.01 GeV^{2}$, adjusted to
$\Delta$ resonances gives better results.

\section{Conclusions.}

The holographical model discussed here allowed us to obtain hadronic spectra with Regge
behavior, not only for the integer spin case, but also for spins 1/2
and 3/2, and also to calculate masses for exotics. In order to do this we considered anomalous dimensions
for operators that create hadrons, and the dilaton that was used has
a form suggested by Einstein's equations, corresponding to
the AdS metric. This two traits allowed the model to reproduce Regge spectra in
all cases considered, and therefore the model can describe hadronic masses
in a unified phenomenological model.


\begin{theacknowledgments}
A. V. acknowledge the financial support from Fondecyt grants 3100028 (Chile).
\end{theacknowledgments}



\bibliographystyle{aipproc}   


\IfFileExists{\jobname.bbl}{}
 {\typeout{}
  \typeout{******************************************}
  \typeout{** Please run "bibtex \jobname" to optain}
  \typeout{** the bibliography and then re-run LaTeX}
  \typeout{** twice to fix the references!}
  \typeout{******************************************}
  \typeout{}
 }



\end{document}